\documentclass[12pt]{article}
\begin{document} 
~~~~~~~~~~~~~~~~~~~~~~~~~~~~~~~~~~~~~~~~~~~~~Preprint INR-TH-2019-002

\begin{center}
{\huge \bf A possibility of CPT violation in the Standard Model } \\[10mm] 
 S.A. Larin \\ [3mm]
 Institute for Nuclear Research of the
 Russian Academy of Sciences,   \\
 60th October Anniversary Prospect 7a,
 Moscow 117312, Russia
\end{center}

\vspace{30mm}
\begin{abstract} 
It is shown that there is a possibility of violation of CPT symmetry
in the Standard Model which does not contradict to the famous CPT theorem.
To check this possibility experimentally it is necessary to increase
the precision of measurements of the proton and antiproton
mass difference
by an order of magnitude. 
\end{abstract}

\newpage
\section{Introduction}
Invariance under the combined transformation $CPT$ is considered
as one of the most fundamental symmetries of local quantum field theory.
Here $C$ is the operator of charge conjugation, $P$ and $T$ are space
 reflection and time reversal operators.
The famous $CPT$-theorem 
was proved in \cite{cpt1}-\cite{cpt2}.
The content of the $CPT$-theorem is approximately as follows:
a Lagrangian of any local Lorentz invariant quantum field theory
with the usual connection between spin and statistics is invariant
with respect to an antiunitary operator $\Theta$ which coincides
with the $CPT$-operator up to a phase (the complete formulation
of the $CPT$-theorem will be given below).

In the present paper we demonstrate that there is a possibility
to violate the $CPT$ symmetry of the Lagrangian of the Standard Model
which does not contradict the $CPT$-theorem.
To check this possibility experimentally it is necessary to increase
the current  precision of measurements of mass difference of a proton 
and an antiproton
$|m_p-m_{\overline{p}}|/m_p$
by an order of magnitude, or it is necessary to increase approximately
twice the precision of the antiproton-to-proton
charge-to-mass ratio experiment \cite{base}.

It should be mentioned that a discovery of violation of such a fundamental
symmetry as the $CPT$ symmetry would be by itself of essential
value. Besides violation of the $CPT$ symmetry may explain the
matter-antimatter asymmetry of the Universe \cite{dol}. 

\section{$CPT$ theorem}
Let us remind the definitions of the operators
of charge conjugation $C$ , 
of space reflection $P$ and time reversal $T$.
 For simplicity
we will consider only fields with spins $0, 1/2$ and $1$. We will use (until
the opposite case is underlined) the interaction representation for fields
which in particular allows to define the  $C$, $P$ and $T$
operators not depending on time even if the corresponding symmetries are violated.

The unitary operator of charge conjugation $C$ acts on  scalar, vector
and spinor fields in the following way.

A scalar field:
\begin{equation}
\label{scalar}
C\phi(x)C^{-1}=\eta_C(\phi)\phi^{+}(x).
\end{equation}

A vector field:
\begin{equation}
CV_{\mu}(x)C^{-1}=\eta_C(V)V_{\mu}^{+}(x).
\end{equation}

A spinor field:
\begin{equation}
\label{spinor}
C\psi(x)C^{-1}=\eta_C(\psi)c\overline{\psi}_T(x),
\end{equation}
where, as usual, $\overline{\psi}=\psi^{+}\gamma_0$ and the unitary
matrix $c$ is defined by equations
\begin{equation}
c\gamma_{\mu}^{T}c^{-1}= -\gamma_{\mu},~~ c^{+}=c^{-1},~~c^{T}=-c, 
\end{equation}
$\gamma_{\mu}$ are Dirac matrices.

 In eqs. (\ref{scalar})-(\ref{spinor}) $\eta_c$ are arbitrary
phase factors. If one demands that the square of the operator $C$ coincides
with the identity, then in the case of Hermitian fields arbirtrariness is 
restricted: $\eta_c$ can have only values $\pm 1$.

The unitary operator of space reflection $P$ is defined as follows.

A scalar field:
\begin{equation}
\label{scalarP}
P\phi(x)P^{-1}=\eta_P(\phi)\phi(x_0,-\overline{x}).
\end{equation}

A vector field:
\begin{equation}
PV_{\mu}(x)P^{-1}=\eta_P(V)\epsilon_{\mu}V_{\mu}(x_0,-\overline{x}),
\epsilon_{\mu}=(1,-1,-1,-1).
\end{equation}

A spinor field:
\begin{equation}
\label{spinorP}
P\psi(x)P^{-1}=\eta_P(\psi)\gamma_0\psi(x_0,-\overline{x}),
\end{equation}

 In eqs. (\ref{scalarP})-(\ref{spinorP}) $\eta_P$ are again arbitrary
phase factors. If one demands that the square of the operator $P$ coincides
with the identity then in the case of boson fields
$\eta_P=\pm 1$ and in the case of fermion fields the phases are restricted
by the values $\pm 1$ or $\pm i$.
  
The antiunitary operator of time reversal $T$ is defined in the following way.

A scalar field:
\begin{equation}
\label{scalarT}
T\phi(x)T^{-1}=\eta_T(\phi)\phi(-x_0,\overline{x}).
\end{equation}

A vector field:
\begin{equation}
TV_{\mu}(x)T^{-1}=\eta_T(V)(-\epsilon_{\mu})V_{\mu}(-x_0,\overline{x}),
~~\epsilon_{\mu}=(1,-1,-1,-1).
\end{equation}

A spinor field:
\begin{equation}
\label{spinorT}
T\psi(x)T^{-1}=\eta_T(\psi)t\psi(-x_0,\overline{x}),
\end{equation}
where in eqs. (\ref{scalarT})-(\ref{spinorT}) $\eta_T$ are
arbitrary phase factors and the unitary matrix $t$ is defined by the condition
\begin{equation}
t\gamma_{\mu}t^{-1}=\gamma_{\mu}^{T}.
\end{equation}

   The operator $T$ can be defined only as an antiunitary operator.
An untiunitary operator has the specific property
\begin{equation}
\Theta \lambda \Theta ^{-1}=\lambda^{*},
\end{equation}
where $\lambda$ is an arbitrary c-number.

Now one can define the
$CPT$ operator which we
denote for shortness as $\Theta$ :
\begin{equation}
\Theta \equiv CPT.
\end{equation}

Transformations for scalar, vector and spinor fields are
\begin{equation}
\label{cpt}
\Theta \phi(x) \Theta^{-1}=\eta_{\Theta}(\phi) \phi^{+}(-x),
\end{equation}

\begin{equation}
\Theta V_{\mu}(x) \Theta^{-1}=\eta_{\Theta}(V) V_{\mu}^{+}(-x),
\end{equation}

\begin{equation}
\label{cpt1}
\Theta \psi(x) \Theta^{-1}=\eta_{\Theta}(\psi)\gamma_5 \psi^{+}_T(-x),
\end{equation}
where $\eta_{\Theta}$ are some phases
and $\gamma_5=i\gamma_0\gamma_1\gamma_2\gamma_3$.
For definiteness one can assume, e.g., the standard representation 
of $\gamma$-matrices. 

It is well known that the operator $\Theta$ can be defined only 
as an antiunitary operator because the time reversal operator $T$
is antiunitary. 

The $CPT$-theorem in the Lagrangian formalism is as follows.
If quantum field theory  satisfies the following six postulates:

1) field equations are local;

2) the Lagrangian is invariant with respect to the proper Lorentz group;

3) one has usual connection between spin and statistics;

4) boson fields commute with all other fields, kinematically
independent fermion fields anticommute;

5) any product of field operators is symmetrized in cases of boson
fields and antysimmetrized in cases of fermion fields (normal
ordering of operators posesses this property);

6) the Lagrangian is Hermitian;

then the Lagrangian of any such theory of interacting fields
with spins 0, 1/2 and 1 is invariant with respect to the following
antiunitary operator $\Theta$ in the Hilbert space:

\begin{equation}
\label{theta}
\Theta \phi(x) \Theta^{-1}= \phi^{+}(-x),
\end{equation}
\[
\Theta V_{\mu}(x) \Theta^{-1}=- V_{\mu}^{+}(-x),
\]
\[
\Theta \psi(x) \Theta^{-1}=-i\gamma_5 \psi^{+}_T(-x),
\]
One can see that this operator $\Theta$ up to phase factors
coincides with the product of three operators $C$, $P$ and $T$
defined in eqs. (\ref{cpt})-(\ref{cpt1}).

Let us now consider the Jost $CPT$-theorem \cite{cpt2} in terms of Wightman
functions (in terms of vacuum
expectations of fields) \cite{sw}.

The postulates of this formalism are: \\
a) invariance of the theory with respect to the proper Lorentz group;\\ 
b) positivity of energy, the existing of vacuum;\\
c) weak causality:
\begin{equation}
<0|\Phi_1(x_1)\Phi_2(x_2)...\Phi_n(x_n)|0>=(-1)^{\sigma}
<0|\Phi_n(x_n)...\Phi_2(x_2)\Phi_1(x_1)|0>
\end{equation}
for all $(x_1,x_2,...x_n)$ for which $\sum_i \lambda_i(x_i-x_{i+1})$
is always a spacelike vector if  $\lambda_i \ge 0$ and $\sum_i \lambda_i=1$,
where $\sigma$ is the number of permutations of fermionic fields.
Weak causality is valid if usual causality conditions of postulates
3) and 4) of the $CPT$-theorem in the Lagrangian formalism are valid,
but it is essentially weaker of these postulates.

The Jost theorem is:  
for any quantum field theory satisfying postulates a)-c),
vacuum expextations are invariant with respect to the operator
$\Theta$ defined in (\ref{theta}),
that is for any set $(x_1,x_2,...x_n)$ one has

\begin{equation}
\label{jost}
<0|\Phi_1(x_1)\Phi_2(x_2)...\Phi_n(x_n)|0>=
\end{equation}
\[
<0|\Theta^{-1}\Theta\Phi_1(x_1)\Theta^{-1}\Theta\Phi_2(x_2)
\Theta^{-1}\Theta...\Theta^{-1}\Theta\Phi_n(x_n)\Theta^{-1}\Theta|0>=
\]
\[
<0|\Theta\Phi_1(x_1)\Theta^{-1}\Theta\Phi_2(x_2)
\Theta^{-1}\Theta...\Theta^{-1}\Theta\Phi_n(x_n)\Theta^{-1}|0>^{*}.
\]

For example, for scalar fields it means
\begin{equation}
<0|\phi_1(x_1)\phi_2(x_2)...\phi_n(x_n)|0>=
<0|\phi_1^{+}(-x_1)\phi_2^{+}(-x_2)...\phi_n^{+}(-x_n)|0>^{*}.
\end{equation}

Let us stress that, in general theory of interacting fields
formulated in terms of  Wightman functions, 
one takes full field operators in the Heisenberg representation.

The check of the equality of masses of particles and antiparticles
is one of fundamental tests of $CPT$ invariance.
Let us consider the case of a proton which, as a stable
particle, is most appropriate for precise direct mass measurements.
The results of experiments are \cite{pdg}
\[
m_p=938.272081 \pm 0.000006~~ MeV ,
\]
\begin{equation}
\label{exp}
|m_p-m_{\overline{p}}|/m_p < 7\times 10^{-10} ~~ at ~~ CL=90\%.
\end{equation}
There are also measurements comparing the charge-to-mass ratio
of the proton and the antiproton \cite{pdg}
\begin{equation}
\label{exp2}
\frac{q_{\overline{p}}}{m_{\overline{p}}}/\frac{q_p}{m_p}=
1.00000000000 \pm 0.00000000007.
\end{equation}
Assuming that the charge of the proton and the antiproton is the same
one gets the accuracy one order of magnitude better then in eq.(\ref{exp}).

The most impressive test of the $CPT$-symmetry comes \cite{pdg} from the limit
on the mass difference between neutral kaons $K^0$ and $\overline{K}^0$ : 

\begin{equation}
|m_{\overline{K}^0}-m_{K^0}|/m_{K^0} \le 0.8\times 10^{-18} ~~ at ~~ CL=90\%.
\end{equation}

One should mention that this restriction is based
on the quantum mechanical picture of neutral kaons as the two-level
system whose evolution is taken in the Wigner-Weisskopf
approximation.

But our further considerations will not concern this special case of neutral
kaon system (and will not contradict to this restriction).

\section{A possibility of $CPT$ violation}

In spite of the theoretical perfectness of the $CPT$-theorem one can
still assume that the operator $\Theta$ defined by this theorem 
in eq.(\ref{theta}) is unphysical, i.e. it does not transform
physical states into physical ones.

One can assume that the physical $CPT$ operator $\Theta_{ph}$ differs
from the theoretical operator $\Theta$ by another choice of the $CPT$
phases $\eta_{\Theta}$ in eq.(\ref{theta}). And it turns out that it 
is possible to violate $CPT$ invariance of the Lagrangian 
of the Standard model
with a non-standard choice of phases $\eta_{\Theta}$ for quark fields.

The Standard Model Lagrangian density is
\begin{equation}
L_{SM}(x)=\frac{g}{2\sqrt{2}} W_{\mu}^{+}(x) \overline{u}\gamma_{\mu}
(1-\gamma_5)\left(d(x)~cos \theta_c+s(x)~ sin\theta_c \right) +h.c. + ...,
\end{equation}
where we have written explicitly only  terms
of interactions of light $u$, $d$ and $s$ quarks with the $W$ bosons
interesting for us.

Here the weak coupling constant $g$ is connected with the Fermi constant
$G_F$ in the usual way:
\begin{equation}
\frac{g^2}{8 M_W^{2}}=\frac{G_F}{\sqrt{2}},~~~~G_F \approx 
\frac{10^{-5}}{m_p^2}.
\end{equation}

The main point is that one can assume that, e.g., $d$ and $s$ quarks
have opposite
$\eta_{\Theta}$ phases with respect to the physical
operator $CPT_{physical} \equiv \Theta_{ph}$ :
\[
\Theta_{ph} d(x) \Theta_{ph}^{-1}=\eta_{\Theta}(d)\gamma_5 d^{+}(-x),~~~~
\Theta_{ph} s(x) \Theta_{ph}^{-1}=\eta_{\Theta}(s)\gamma_5 s^{+}(-x),
\]
\begin{equation}
\label{min}
\eta_{\Theta}(s)=-\eta_{\Theta}(d),
\end{equation}
where $\eta_{\Theta}(d)=-i$, as in eq.(\ref{theta}).
It can be arranged, e.g., in  models of composite quarks. 

To see this let us suppose that 
fields corresponding to $d$ and $s$ quarks are composed
of a constituent spinor field $\psi$
in a model of composite quarks in the following way:
\begin{equation}
\label{composed}
d=(\overline{\psi}\gamma_0 \psi) \psi,
\end{equation}
\[
s=(\overline{\psi}\gamma_5 \psi) \psi,
\]
where normal ordering of operators is assumed.

Now one can take into account that
\begin{equation}
CPT\overline{\psi} O\psi(CPT)^{-1}=
\eta(\psi)^2\overline{\psi}\gamma_5\gamma_0 O^{+}\gamma_0\gamma_5\psi,
\end{equation}
where $O$ is some combination of $\gamma$-matrices
and $\eta(\psi)$ is the CPT-phase of the constituent field $\psi$.
Hence
\begin{equation}
\label{bilinear}
CPT\overline{\psi} \gamma_0\psi(CPT)^{-1}=
-\eta(\psi)^2\overline{\psi}\gamma_0\psi,
\end{equation}
\[
CPT\overline{\psi} \gamma_5\psi(CPT)^{-1}=
-\eta(\psi)^2\overline{\psi}\gamma_5\psi.
\]
Using eqs. (\ref{bilinear}) one gets the following result
for $CPT$ transformations of $d$ and $s$ quarks
\begin{equation}
\label{compositeds}
CPT d(x)(CPT)^{-1}=\eta(\psi)^3 \gamma_5 d^{+}_T(-x),
\end{equation}
\[
CPT s(x)(CPT)^{-1}=-\eta(\psi)^3 \gamma_5 s^{+}_T(-x).
\]
Thus  $d$ and $s$ quarks obtain opposite $CPT$ phases in 
such a model of composite quarks.

In this case the Standard Model Lagrangian will not be any more 
invariant under the physical
$CPT$ operator $ \Theta_{ph}$ but will consist of the $CPT$ even and
$CPT$ odd parts.

At first glance 
this extra minus in eq.(\ref{min}) contradicts to the powerful
Jost theorem which does not allow extra minuses in 
eqs.(\ref{jost}),(\ref{theta}).
But one should remember that the fields there are the full operators in the Heisenberg
representation. The operator $ \Theta_{ph}$ does not commute with
the Hamiltonian, hence it depends on time and is not restricted by
the Jost theorem.

We would like to underline that the operator $ \Theta_{ph}$ does not commute with
the Lagrangian both in the interaction representation and in the  Heisenberg
representation. We refer here to the  Heisenberg
representation just to demonstrate that our choice of opposite $CPT$-phases
for $d$ and $s$ quarks does not contradict the Jost theorem.

We underline once more that the considered type of $CPT$ violation appears because we can choose
the opposite $CPT$ phases for $d$ and $s$ quarks. Then the terms
of the weak Lagrangian, which are linear in $d$ and $s$
fields, obtain different signs after applying the $CPT$ operator. This type of
$CPT$ violation  happens
in a theory with an interaction Lagrangian containing 
at least two terms with odd number of fields of different
$CPT$-phases.

Let us consider the influence of this $CPT$ violation on the difference
of proton and antiproton masses. For this purpose we will use the simplified
quantum mechanical picture of a proton as a superposition
\begin{equation}
|p>=|uud>+\xi |uus>,
\end{equation}
where $|uud>$ and $|uus>$ are the eigenvectors of the Hamiltonian
of strong interactions consisting of the corresponding light quarks $u$, $d$ and $s$.

The amplitude $\xi$ appears after the mixing of these states
due to the Hamiltonian of weak interactions $H_w$:
\begin{equation}
\xi \approx \frac{<uud|H_w|uus>}{m_{\Sigma}-m_p} 
\approx \frac{G_F sin\theta_c}{m_{\Sigma}-m_p}\approx 0.8 \times 10^{-5},
\end{equation}
here $m_{\Sigma}\approx 1189~MeV$, the $\Sigma ^{+}$-hyperon mass.

One can assume that the physical $CPT$-operator $\Theta_{ph}$ transforms
the  proton state into the physical antiproton state
\begin{equation}
\label{antiprotonstate}
\Theta_{ph}|p>=\eta_p\left(|\overline{u}~\overline{u}~\overline{d}>-\xi^*
 |\overline{u}~\overline{u}~\overline{s}>\right)=\eta_p|\overline{p}>
\end{equation}
where  $\eta_p$ is the proton $CPT$-phase factor, the minus sign appears
due to opposite $CPT$-phases of $d$ and $s$  quarks and $|\overline{p}>$ is the physical
antiproton state.

Correspondingly the action of $\Theta_{ph}$ on the antiproton state will give the proton 
state
\begin{equation}
\Theta_{ph}|\overline{p}>=\eta_{\overline{p}}|p>.
\end{equation}

This case is similar to the case of the standard charge conjugation operator $C$.
The operator $C$ is assumed to transform physical particles to physical antiparticles
although the Lagrangian of weak interactions is not invariant with respect to $C$.

On the other hand the unphysical operator  $\Theta$ will act on the proton state
as follows:
\begin{equation}
\label{unphysicalstate}
\Theta|p>=\eta_p\left(|\overline{u}~\overline{u}~\overline{d}>+\xi^*
 |\overline{u}~\overline{u}~\overline{s}>\right),
\end{equation}
producing an unphysical state which differs from the physical antiproton state
(\ref{antiprotonstate}) only by the phase in front of $\xi^*$.

Let us remind that the minus sign in front of $\xi^*$ 
for the physical untiproton in eq.(\ref{antiprotonstate})
was obtained due to the opposite $CPT$ phases of $d$ and $S$ quarks
which can be arranged, e.g., in models of composite quarks, see eq.(\ref{compositeds}).

The full Hamiltonian is the sum of the $CPT$ even and $CPT$ odd parts:
\begin{equation}
H=H_{+}+H_{-}.
\end{equation} 

The masses of the proton and the antiproton are
\begin{equation}
\label{pm}
m_p=<p|H_{+}|p>+<p|H_{-}|p>,
\end{equation}
\[
m_{\overline{p}}=<\overline{p}|H_{+}|\overline{p}>+
<\overline{p}|H_{-}|\overline{p}>.
\]

Applying the $CPT$- operator $\Theta_{ph}$ we get
\begin{equation}
\label{apm}
m_{\overline{p}}=<\overline{p}|\Theta_{ph}^{-1}\Theta_{ph}H
\Theta_{ph}^{-1}\Theta_{ph}|\overline{p}>=<p|H_{+}|p>-<p|H_{-}|p>.
\end{equation}

Subtracting (\ref{pm}) and (\ref{apm}) one has
\[
m_p-m_{\overline{p}}=2<p|H_{-}|p>=
\]
\[2<uud+\xi~ uus|H_{-}|uud+\xi~uus>
\approx 4\xi <uud|H_{-}|uus>,
\]
\begin{equation}
|m_p-m_{\overline{p}}|/m_p
\approx 4\xi sin\theta_c G_F \approx 6 \times 10^{-11},
\end{equation}
which should be compared with present experiments, 
see eqs. (\ref{exp}),(\ref{exp2}).

Thus to check the possibility of $CPT$ violation
it is necessary to improve the current experimental precision
for $|m_p-m_{\overline{p}}|/m_p$ approximately by an order of magnitude.
It concerns direct measurements of $|m_p-m_{\overline{p}}|/m_p$. 

If one considers the charge-to-mass ratio mesurements
and assumes that the charge of the proton and the antiproton is the same then 
the precision is already one order of magnitude better $(7\times 10^{-11})$,
see eq.(\ref{exp2}). In this case it is desirable to improve the precision
of experiments approximately twice.

There are of course other well known consequences of $CPT$-invariance which can be checked
experimentally. For example cross-sections of scattering processes of some particles should
coincide with cross-sections of reversed reactions of corresponding antiparticles with opposite spins.
More precisely the transition probabilities for the following two processes should coincide:
1) particles with momenta \boldmath{$p_i$} and spins $\sigma_i$ react to produce particles
with momenta \boldmath{$p'_i$} and spins $\sigma'_i$; 2) the corresponding antiparticles
with \boldmath{$p'_i$},-$\sigma'_i$ react to produce antiparticles with \boldmath{$p_i$}
and -$\sigma_i$.

 Also there are interesting checks of $CPT$ invariance in the system of neutral kaons, see e.g.
reviews in \cite{pdg}.
But the simplest tests of $CPT$ invariance are the tests of equality of masses and life times
of particles and their antiparticles. The measurement of the proton and antiproton mass difference
seems to be the most proper experiment to check the possibility of violation of the $CPT$ symmetry
considered in the present paper.

\section{Conclusions}
We have shown that there is still the possibility of violation of $CPT$
symmetry in the Standard Model. It can be achived by the proper
choice of the $CPT$-phases for quarks. 
To check this possibility experimentally it is necessary to increase
the accuracy of direct measurements of the proton and antiproton mass difference
by an order of magnitude or to increase twice the accuracy of the charge-to-mass
ratio experiment.

\section{Acknowledgements}
The author is grateful to the collaborators of the Theory division of INR
for helpful discussions.


\newpage

\begin{thebibliography}{99}
\bibitem{cpt1} G. Lueders, Dan.Mat.Fys.Medd. 28 No5 (1954).
\bibitem{pa}W. Pauli, In: 'Niels Bohr and the Development of Physics',
 McCraw-Hill, New-York and Pergamon Press, London (1955).
\bibitem{lu}G. Lueders, Ann. Phys. 2 (1957) 1.
\bibitem{sc}J. Schwinger, Phys. Rev. 82 (1951) 914, ibid. 91 (1951) 713.
\bibitem{be}J.C. Bell, Proc.Roy.Soc.Lond. A 231 (1955) 479.
\bibitem{cpt2} R. Jost, Helv.Phys.Acta 30 (1957) 409.
\bibitem{base} BASE Collaboration (S. Ulmer  et al.),
Nature 524 (2015) 196.
\bibitem{dol} A.D. Dolgov, Phys.Atom.Nucl. 73 (2010) 588.
e-Print: arXiv:0903.4318 [hep-ph].
\bibitem{sw} R.F. Streater and A.S. Wightman. PCT, Spin and Statistics
and All That. W.A. Benjamin, INC New York - Amsterdam, 1964. 
\bibitem{pdg} M. Tanabashi et al. (Particle Data Group), 
Phys.Rev. D98 (2018) 030001. 
\end{thebibliography}
\end{document}